\documentclass[10pt,letterpaper]{article}
\usepackage[top=0.85in,left=2.75in,footskip=0.75in]{geometry}

\usepackage{amsmath,amssymb}

\usepackage{changepage}

\usepackage[utf8x]{inputenc}

\usepackage{textcomp,marvosym}

\usepackage{cite}

\usepackage{nameref,hyperref}

\usepackage[right]{lineno}

\usepackage{microtype}
\DisableLigatures[f]{encoding = *, family = * }

\usepackage[table]{xcolor}

\usepackage{array}

\newcolumntype{+}{!{\vrule width 2pt}}

\newlength\savedwidth

\raggedright
\usepackage[aboveskip=1pt,font=bf,labelsep=period,justification=raggedright,singlelinecheck=off]{caption}

\makeatletter
\renewcommand{\@biblabel}[1]{\quad#1.}
\makeatother

\usepackage{lastpage,fancyhdr,graphicx}
\usepackage{epstopdf}
\fancyheadoffset[L]{2.25in}
\fancyfootoffset[L]{2.25in}
\usepackage{booktabs}
\usepackage{xspace}
\usepackage{tabularx}
\usepackage{multirow}
\usepackage{subcaption}
\usepackage{colortbl}

\definecolor{light-gray}{gray}{0.95}

\newcommand{\BDCT}{BDCT\xspace} %Bluetooth-based decentralized contact tracing

\begin{document}
\vspace*{0.2in}

\begin{flushleft}
{\Large
\textbf\newline{An Empirical Evaluation of Bluetooth-based Decentralized Contact Tracing in Crowds} % Please use "sentence case" for title and headings (capitalize only the first word in a title (or heading), the first word in a subtitle (or subheading), and any proper nouns).
}
\newline
\\
Hsu-Chun Hsiao\textsuperscript{1,2},
Chun-Ying Huang\textsuperscript{3},
Shin-Ming Cheng\textsuperscript{2,4},
Bing-Kai Hong\textsuperscript{4},
Hsin-Yuan Hu\textsuperscript{5},
Chia-Chien Wu\textsuperscript{1},
Jian-Sin Lee\textsuperscript{5},
Shih-Hung Wang\textsuperscript{6},
Wei Jeng\textsuperscript{5*}
\\
\bigskip
\textbf{1} Department of Computer Science and Information Engineering, National Taiwan University, Taiwan
\\
\textbf{2} Research Center for Information Technology Innovation, Academia Sinica, Taiwan
\\
\textbf{3} Department of Computer Science, National Yang Ming Chiao Tung University, Taiwan
\\
\textbf{4} Department of Computer Science and Information Engineering, National Taiwan University of Science and Technology, Taiwan
\\
\textbf{5} Department of Library and Information Science, National Taiwan University, Taiwan
\\
\textbf{6} Information Networking Institute, Carnegie Mellon University, USA
\\
\bigskip

% Insert additional author notes using the symbols described below. Insert symbol callouts after author names as necessary.
% 
% Use the asterisk to denote corresponding authorship and provide email address in note below.
* wjeng@ntu.edu.tw

\end{flushleft}
\section*{Abstract}
Many countries are using \emph{digital contact tracing} to help contain COVID-19’s spread. Among various available techniques, decentralized contact tracing that uses Bluetooth received signal strength indication (RSSI) to detect proximity is considered a smaller privacy risk than approaches that rely on collecting absolute locations via GPS, cellular tower history, or QR-code scanning. % As of October 2020, 
Approximately half a year after WHO declared COVID-19 a pandemic, there have been millions of downloads of such Bluetooth-based contact-tracing apps as more countries officially adopt them. However, the effectiveness of these apps in the real world remains unclear due to a lack of empirical research that includes realistic crowd sizes and densities. This study aims to fill that gap by empirically investigating the effectiveness of Bluetooth-based decentralized contact tracing (\BDCT) in crowd environments with a total of 80 participants, emulating classrooms, moving lines, and other types of real-world gatherings. The results confirm that Bluetooth RSSI is unreliable for detecting proximity, and that this inaccuracy worsens in especially crowded environments. In other words, this technique may be least useful when it is most needed, and it is fragile when confronted by low-cost jamming. Moreover, technical problems such as high energy consumption and phone overheating caused by the contact-tracing app were found to negatively influence users’ willingness to adopt it. On the bright side, however, Bluetooth RSSI may still be useful for detecting coarse-grained contact events; for example, proximity of up to 20m lasting one hour. Based on our findings, we recommend that existing \BDCT apps can be repurposed to focus on coarse-grained proximity detection, and future apps may calibrate distance estimates and adjust broadcast frequencies based on auxiliary information.

%\section*{Author summary}

%\linenumbers

\section*{Introduction}

\emph{Contact tracing} has been known to be an effective method for controlling the spread of infectious diseases. In the traditional contact-tracing model, trained personnel use in-person or telephone interviews to identify and list those who have had meaningful exposure to diagnosed individuals during a disease’s likely transmission period~\cite{lee2021benefits}. The risk levels of such contact can then be determined based on factors such as the physical distance between healthy and infected people, the duration of such exposure, their mobility trajectories over time, whether masks were worn, and whether the contact occurred indoors or outdoors.

With the escalating scale of the COVID-19, the vast numbers of new cases identified via testing each day exceed the capacity of traditional contact tracing, the failures of which have been extensively reported~\cite{lo2021manual,shahroz2021manual}. Some local health services have even abandoned their former practice of directly communicating with the close contacts of a case~\cite{lewis2020manual}. Thus, governments are turning to \emph{digital contact tracing} that utilizes mobile providers or mobile applications to help identify potential contacts by means of GPS data, cell tower connection history, or QR-code scanning.

However, large-scale collection of citizens’ digital footprints by the government has sparked concerns about mass surveillance~\cite{maytin2021privacy,romero2021privacy}. Accordingly, to enhance the privacy of digital contact tracing, a number of decentralized approaches have been developed, which are capable of detecting close proximity between phones (and thus, presumably, those phones’ owners) without needing to report each phone’s location to a centralized server. 
Many use Bluetooth received signal strength indication (RSSI) as a proxy for distance and share the following high-level procedures~\cite{ExposureNotification,covidwatch}:

\begin{enumerate}
\item The app broadcasts Bluetooth tokens periodically (e.g., every 100-270 ms). The tokens are anonymized and changed from time to time (e.g., every 15 minutes) to balance privacy and storage overhead. %These tokens contain no information about the phone or its owner. 
\item Each phone stores both sent and received tokens.
\item If a user has tested positive, she can voluntarily publish all the tokens her phone sent during her potential transmission period (e.g., the past 14 days). This upload process may require authorization from a health authority to prevent false reporting.
\item The app downloads tokens published by infected people and compares them with those received locally. The app then uses the overlapping subset (i.e., those sent from infected people and received by the user), with their timestamps and RSSI values, to estimate distance and exposure duration.
 \item The app may have built-in risk-estimation models and recommend different measures (e.g., self-monitoring, commencing quarantine, or seeking medical attention) based on the estimate risk levels. The user can then voluntarily seek help from health authorities if the exposure risk is considered high.
 \end{enumerate}
 
As countries lift their lockdowns and reopen facilities, there is a rush to deploy these Bluetooth-based decentralized contact-tracing apps. As of January 2021, at least 38 countries and 24 U.S. states have adopted these apps~\cite{mosoff2020countries,rahman2021countries}. The EU has been working on a Europe-wide coronavirus-tracing network, based on new infrastructure that will enable data-sharing and ensure interoperability between national contact-tracing apps~\cite{noll2020eu,eu2020gateway}. Apple and Google jointly released an exposure notification (GAEN) API and Exposure Notification Express~\cite{GAENBluetoothSpec}, helping local health authorities develop their own apps.

Nevertheless, few studies have gauged the effectiveness of these apps in practice, with existing evaluations of Bluetooth-based decentralized contact-tracing (\BDCT) techniques relying either on simulations with mathematical models, or experiments involving only mobile devices or a very limited number of participants. Existing evaluations of Bluetooth-based indoor positioning~\cite{zafari2019survey} are not directly applicable to contact tracing due to different settings and assumptions. For example, indoor positioning often assumes using trilateration or known floor maps and considers only one or a few sending devices within the communication range.

Accordingly, this study empirically investigates the effectiveness of \BDCT apps in two phases. In Phase 1, we will examine whether Bluetooth RSSI can reliably predict distance in a controlled experimental setting. In Phase 2, based on estimated distances between pairs of participants’ phones over time, we compare detected proximity and contact events in a semi-controlled event: a real-world academic gathering, the ground truth of which will be carefully recorded. We recruited a total of 80 participants to use one Bluetooth-based contact-tracing app, which we modified from Covid-Watch-TCN~\cite{covidwatch}, in our controlled and semi-controlled settings. Our modifications allowed us to collect ground-truth data and the phones’ usage logs. After both experimental phases were completed, we also conducted a follow-up survey with participants to enrich the data we had already obtained.

This paper will address the following questions:
%This paper seeks to address the following questions:
 \begin{itemize}
 \item RQ1. Can the app reliably estimate the distance between its users based on Bluetooth RSSI under different crowd parameters, e.g., standing still vs. walking, with or without physical barriers, varying interpersonal distances, and the presence of jamming?
 %RQ1. Can a Bluetooth-based contact tracing application reliably estimate the distance based on Bluetooth RSSI under different parameters in the crowd environments i.e., a) the mobile OS, b) still/walking, c) with or without physical barriers, d) the distance, and e) the presence of jamming sources? 
 % How well and to what degree does a Bluetooth-based contact-tracing application detect proximity events in crowd environments, compared with the actual exposure (the ground truth), in terms of the different factors i.e., a) the mobile OS, b) still/walking, c) physical barriers, and d) the distance?
 \item RQ2. How accurately does the app detect proximity and contact events in realistic crowd environments, as compared with the ground truth of such events?
 % RQ2. How well does a Bluetooth-based contact-tracing application accurately detect proximity and contact events in real-world crowd environments, compared with the actual exposure (the ground truth)? 
 % Given the RQ1, how would the accuracy of the proximity detection be changed and influenced by malicious attacks?
  \item RQ3. What are users’ perceptions and experiences of using these apps?
  % RQ3. What is users' perception and experience on using these apps?
 \end{itemize}

Our experimental results suggest that Bluetooth RSSI is unreliable for detecting proximity, and reveals that such inaccuracy worsens in crowded environments. This implies that this technique may be least useful when it is most needed, and fragile when confronted by low-cost jamming. Specifically, the app failed to capture the majority of proximity events: only 16 out of 67 (24\%) proximity events were detected by the app when setting a 2-meter distance threshold; only 19 out of 67 (28\%) were detected even when no distance threshold was set. In terms of user experience and perceptions, technical problems such as high energy consumption and phone overheating caused by the app were found to negatively influence users’ willingness to adopt it. On the bright side, the app captured 63\% of contacts lasting one hour in a room containing 50 participants and more than 150 other people. Divided by operating system, 80\% of Android devices were able to be discovered by both nearby Android and iOS devices in about an hour. This implies that this technique may still be useful for detecting coarse-grained contact events; for example, contacts within 20m that last for at least an hour. The sampled phone users said they were more willing to use the similar apps 1) when in crowded environments and 2) while contact tracers from health departments were also using them.

Although our study is based on a specific implementation, many of our instruments and findings are applicable to \BDCT applications~\cite{shubina2020survey} with similar designs. We discuss limitations in the Practical Implications section.

\section*{Background and Related Work}
\label{sec:related}

We summarize several representative privacy-preserving contact-tracing methods and previous studies that review them.

\subsection*{Privacy-preserving contact tracing for COVID-19}
There are two major approaches to privacy-preserving contact tracing. The first adopts a decentralized design intended to minimize the amount of data needed to be sent to a centralized server. The other applies cryptographic algorithms to protect sensitive user data. 

\paragraph{Decentralized design} This approach is exemplified by Safe Path~\cite{raskar2020apps}, DP3T~\cite{troncoso2020decentralized}, and Covid-Watch~\cite{covidwatch}. Safe Path logs a user’s movement routes in his/her mobile device and only exports that data to health authorities if that user is diagnosed with the virus. When this happens, the exported dataset is first redacted to ensure privacy, and then is broadcast to other users to allow them to self-determine their likelihood of having been exposed. Rather than capturing and storing route information, DP3T and Covid-Watch continuously broadcast anonymized tokens, which will only be published in the event of a positive diagnosis. This enhances privacy because everyone can locally infer exposures based on whether any received tokens belong to infected people.

\paragraph{Cryptographic algorithms} This approach has been adopted by Private Set Intersection (PSI)~\cite{berke2020assessing} and TraceSecure~\cite{bell2020tracesecure}. PSI enables two parties to compute the intersection of their data in a privacy-preserving way, with only the common data values being revealed. The data used for computation contains only hashed location points, and location privacy can therefore be guaranteed. TraceSecure, on the other hand, incorporates a public-key-based security protocol for message exchange and storage. An optional homomorphic encryption scheme can be used to further enhance data protection. While this cryptographic-based approach guarantees better privacy than the decentralized one, it is less deployable on consumer devices and existing infrastructure. 

Our study therefore focuses on decentralized contact tracing, since it is more feasible to deploy on the massive scale required for this purpose.

\paragraph{Implementation} Among the decentralized contact-tracing apps, a number of them use Bluetooth RSSI as a proxy for close proximity between devices, and therefore, between the owners of those devices. Early implementations such as the DP3T project and Covid-Watch had to confront Bluetooth interoperability issues between the two major mobile platforms, iOS and Android. In April 2020, the two major smartphone OS providers, Apple and Google, announced that they had formed a coalition to release APIs that would help contact-tracing apps work across iOS and Android devices, with the first such APIs appearing the following month~\cite{GAENBluetoothSpec}. Android version 6.0 and iOS 13.7 and higher have supported the fundamental functions of Bluetooth contact tracing, including broadcasting and listening to tokens, at the OS level. To prevent abuse of these APIs, the companies restricted each country to one official app. An increasing number of national projects have adopted the GAEN APIs, including Switzerland’s SwissCOVID, Italy’s Immuni, and Germany’s Corona-Warn-App. Since our use of it in our experiments, Covid-Watch has also now switched to using the GAEN APIs. We plan to use our proposed methodology to evaluate GAEN-API-based apps once the GAEN APIs are enabled in Taiwan.

\paragraph{Possible DoS attacks} Studies have pointed out that current contact tracing apps, based on ephemeral IDs, are vulnerable to DoS attacks. Chen and Hu~\cite{chen2020mitigating} presented BlindSignedIDs, which are verifiable ephemeral identifiers, using blind signatures and TESLA authenticators to verify EphIDs in-place. BlindSignedIDs reduce storage requirements by more than 90\% and are demonstrated effective in mitigating gigabyte-level DoS attacks.

\paragraph{Privacy concerns} Tracing contacts by accessing devices’ relative locations (e.g., using Bluetooth signal reception) is considered more protective of user privacy than capturing absolute locations (e.g., via GPS)~\cite{ahmed2020relative,alsahli2021relative}. Several prior studies have investigated privacy threats such as replay attacks and de-anonymization attacks by state-level or resourceful adversaries~\cite{cho2020contact, baumgartner2020mind}, or have proposed advanced cryptographic solutions to enhance privacy~\cite{trieu2020epione}. While detailed privacy concerns are beyond the scope of our empirical study, we feel it should be noted that privacy enhancement beyond a certain level is likely to degrade the detection accuracy and other aspects of the performance of contact-tracing apps. Our methodology and protocols used in this study will be useful in assessing whether the negative performance impacts of future privacy-enhancement efforts outweigh their benefits.

\subsection*{Evaluation of Bluetooth-based Contact Tracing}
Broadcasting of anonymized tokens to nearby devices is fundamental to the implementation of privacy-preserving contact tracing. In contemporary smartphones, the most widely deployed techniques that support such broadcasting are Bluetooth and Wi-Fi. Of the two, however, developers prefer to use Bluetooth because it was originally designed to function within ad hoc networks, and because it has also been widely used for distance measurement and indoor positioning~\cite{BCCGD16,guo2020hybrid}. By reading an RSSI reported by a receiver, an application can estimate the distance between the receiving and sending devices; and indoor positions can be calculated based on three or more RSSIs from fixed-location broadcasters or beacons.

\paragraph{RSSI-based distance estimation} Bluetooth technology was developed to replace cables connected to peripherals. Its lightweight design and wide deployment enable many IoT and logistic applications, such as warehouse management and traffic monitoring~\cite{song2021survey}. Bluetooth's RSSI can also help distance measurement and indoor positioning~\cite{zafari2019survey}. A receiving device can estimate its distance from a sending device based on the perceived RSSI. Researchers also attempt to use alternative radio-frequency-based techniques such as Zigbee, Ultra-Wideband (UWB), and WiFi for positioning and contact tracing~\cite{bian2020zigbee,trivedi2021wifi}. However, Bluetooth remains the mainstream choice because of its cost, efficiency, and availability. The discussion of other alternatives is therefore outside the scope of our study.

A major drawback to RSSI-based distance estimation is the variation in its measurement results, caused by various environmental factors such as interference, emission power, and receiver sensitivity, all of which introduce noise. Indoor positioning applications have improved estimation accuracy through trilateration (using multiple referenced sending devices at known positions), incorporating floor maps, or training position-dependent signal attenuation models. Some have proposed augmenting Bluetooth with other sensors to improve accuracy~\cite{guo2020hybrid}. However, these improvements may be difficult to apply to \BDCT because they will require national-scale referenced device deployment, indoor mapping, or modeling. Moreover, \BDCT and other Bluetooth-based applications consider different settings. For example, in \BDCT, every user's phone is both sending and receiving, thus more likely to saturate wireless channels than indoor positioning (which requires only one or a few sending devices present in a room). Therefore, a thorough evaluation is required to understand \BDCT's limitations and possibilities.

Nevertheless, we felt that Bluetooth-based distance estimation has the potential to provide helpful information to pandemic investigators, and tested its performance for this purpose with human subjects in controlled and uncontrolled settings, as explained in the Research Design section~\ref{sec:methodology}.

\paragraph{Empirical evaluation of contact tracing} Although many \BDCT apps have been deployed in the field, their effectiveness remains unknown due to the lack of ground-truth information. Prior to our present effort to help fill that gap, an empirical study~\cite{jackie2020proximity} was conducted in April 2020 among a group of 48 soldiers in Germany. They were divided into five scenarios with different moving patterns, with at most 10 people in any one scenario. A follow-up report documented the experimental protocol and provided some preliminary analysis, but drew no clear conclusions and made no recommendations.

Studies~\cite{opentrace-calibration,leith2020coronavirus} have shown that distances derived from RSSIs without calibration can be quite diverse, even in controlled-experiment scenarios with no human participants and with the same settings on all devices. 
Leith and Farrell conducted a series of studies~\cite{leith2020measurement,leith2021measurement,leith2020coronavirus} to empirically measure RSSI between mobile phones indoors and outdoors, as well as on a bus and a tram, and considered factors that could affect such signal strength, including distance, phone orientation, and absorption and/or reflection by surroundings such as building walls or even human bodies. Their follow-up measurement study recruited five participants on a commuter bus to investigate the relationship between Bluetooth attenuation and distance in an environment prone to signal reflection. They made several recommendations for improving \BDCT's, such as leaving phones on tables instead of keeping them in bags or pockets.

Our study considers scenarios involving much larger groups of participants and jamming devices, which allows us to simulate crowded scenarios and observe issues that might occur only or mostly in large groups, e.g., rapid battery depletion, interference, and interoperability problems between different phone models. In addition, we also collected users’ feedback after using a Bluetooth-based contact-tracing app, particularly their perceptions and concerns regarding privacy and usability.

Since our focus is on Bluetooth-related issues, the following studies were also important to our thinking, despite being beyond the scope of our own research.

Existing exposure-notification apps often feature fixed thresholds for identifying contact events and calculating exposure risks. For example, if a user has been in close proximity with a confirmed patient for a sufficiently long time (e.g., 15 minutes), that user will be warned of potential exposure by the app. Wilson et al.~\cite{wilson2021quantifying} proposed a calibrated measure of infection risk based on empirical measurements, and devised a risk-scoring system that aims to provide better quarantine recommendations.

Some scholars have evaluated the effectiveness of contact tracing via mathematical modeling and simulation, and compared it against other countermeasures such as social distancing or lockdowns~\cite{kucharski2020effectiveness, abueg2021modeling}. Our findings can provide more realistic parameters for contact tracing that can assist the refinement of such models and simulations.

\subsection*{Worldwide Government Practices and User Perceptions}
Despite the security and privacy issues involved in the adoption of digital contact tracing, as of October 2020 (i.e., approximately half a year after WHO declared COVID-19 a pandemic), a considerable number of national and local governments have either already introduced such measures in the fight against COVID-19, or are planning to do so~\cite{perrigo2020countries}.

On the flipside, a relatively small number of governments have launched new human-based tracing services or announced improvements to existing ones~\cite{lee2020manual}, and remain on the fence regarding the adoption of digital contact tracing, repeatedly citing concerns about uptake rates and false positives/negatives~\cite{hern2020uptake,vergallo2021false}. The UK, for instance, planned to launch a coronavirus app that enables users to report symptoms and book tests, but does not allow contact tracing~\cite{allison2020uk}. In Canada, Manitoba's chief public health officer declared that the contact-tracing app ``will not replace public health’s ability to contact trace'', although the other four provinces have adopted such an app~\cite{mcguckin2020manitoba}.

Several large-scale questionnaire surveys have been conducted to capture phone users’ general perceptions toward contact-tracing mobile apps. A multi-country survey of Europe and North America has shown high user acceptance of downloading such apps (74.8\%)~\cite{altmann2020acceptability-pub}. However, the results of another survey, conducted in the U.S., suggest that support for the policy of encouraging use of these apps is relatively weak (42\%), as compared to traditional measures; but also that the implementation of decentralized data storage helps increase acceptance~\cite{zhang2020americans}. According to a UK survey, 67.2\% of those who are unwilling to participate in app-based contact tracing considered privacy concerns the main reason~\cite{bachtiger2020uk}. A team in Jordan, meanwhile, reported that 71.6\% of their respondents accepted the use of contact-tracing technology, but only 37.8\% actually used it~\cite{abuhammad2020acceptability}. Additionally, a German team found that factors such as age, gender, education, and income could influence the download and use of contact-tracing apps~\cite{grill2021germany}. It is noteworthy, therefore, that all of the respondents to our post-study survey had actually experienced using such an app, and were thus able to provide meaningful, app-specific answers about their usage experience and  privacy perceptions.
\section*{Research Design}
\label{sec:methodology}

\begin{table*}[t]
    \centering
    \caption{Summary of our methodology}
    \label{tab:methodology}
    %\begin{tabularx}{\textwidth}{|c|p{2cm}|p{2.3cm}|X|p{2.8cm}|}
    \begin{tabularx}{\textwidth}{|c|p{1.9cm}|p{1.8cm}|X|p{2.5cm}|}
        \hline
         \textbf{RQ} & \textbf{Element} & \textbf{Sample} & \textbf{Description} & \textbf{Collected Data} \\ \hline
         N/A & App customization and configuration & N/A 
         & The app was built on the source code of the Covid-Watch Exposure Notification App. Device information was captured when a participant registered as a user.
         & \textbullet Operating system \newline 
         \textbullet Phone model \\ \hline
         RQ1
         & Controlled experiment 
         & 30 participants: \newline
         14 Android devices; \newline
         16 iOS devices
         & The experiment was conducted in both indoor and outdoor scenarios. Participants held their devices and followed instructions to stand still, walk along assigned paths, move closer to one another, etc.
         & 
         \textbullet Timestamp \newline
         \textbullet Battery status \newline
         \textbullet Sent/Received tokens \newline
         \textbullet Unique device IDs
         \\ \cline{1-4}
         RQ1 & Semi-controlled experiment
         & 50 participants: \newline 
         24 Android devices; \newline
         26 iOS devices
         & The experiment was performed during a conference in an auditorium. Participants carried their devices when they were listening to a speech, having a group discussion, taking a tea break, etc. 
         & 
         \textbullet RSSI values \newline
         \textbullet Calculated distances \newline
         \textbullet Ground-truth records
         \\ \hline
         RQ3 & User survey & 24 respondents
         & A follow-up survey was developed to reveal the participants’ 1) technical problems encountered during both experiments, 2) willingness to use contact-tracing apps, and 3) privacy concerns.
         & \textbullet Survey results
         \\ \hline
    \end{tabularx}
\end{table*}

\begin{figure*}[t]
     \centering
     \begin{subfigure}[b]{0.3\textwidth}
         \centering
         \includegraphics[width=\textwidth]{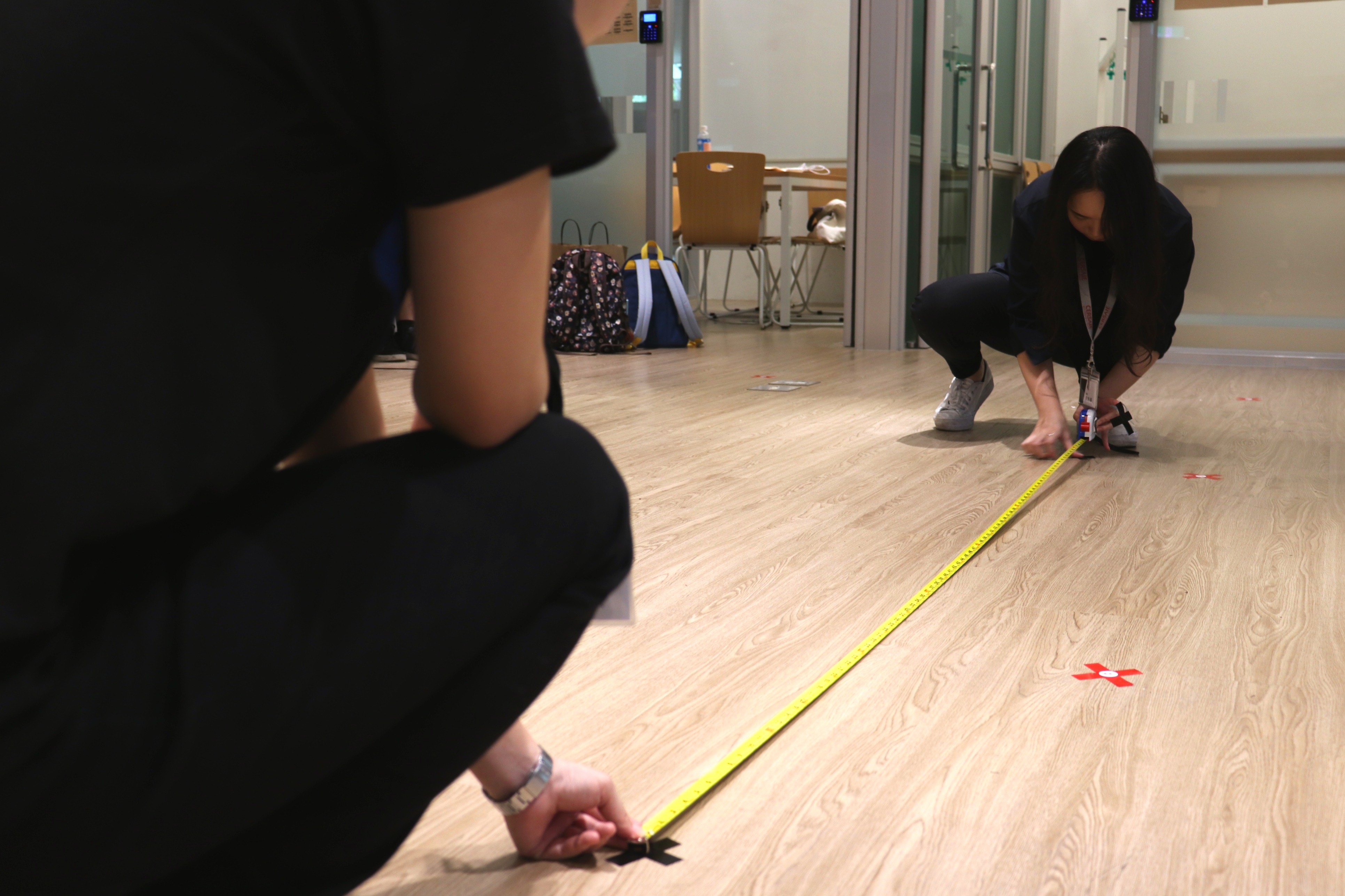}
         \caption{The floor marked with tapes}
         \label{fig:exp_pic_1}
     \end{subfigure}
     \hfill
     \begin{subfigure}[b]{0.3\textwidth}
         \centering
         \includegraphics[width=\textwidth]{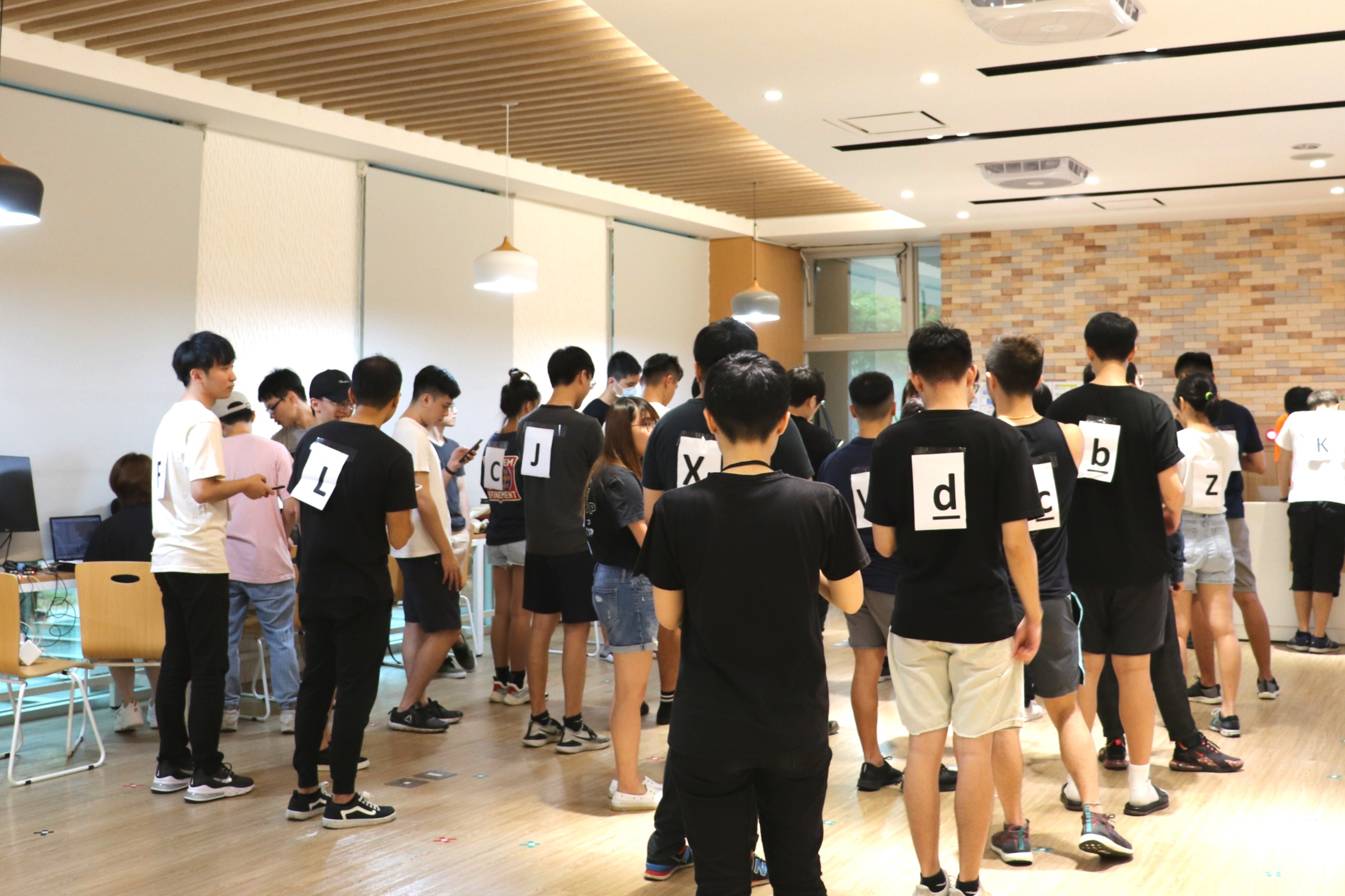}
         \caption{The indoor scenario (a classroom)}
         \label{fig:exp_pic_2}
     \end{subfigure}
     \hfill
     \begin{subfigure}[b]{0.3\textwidth}
         \centering
         \includegraphics[width=\textwidth]{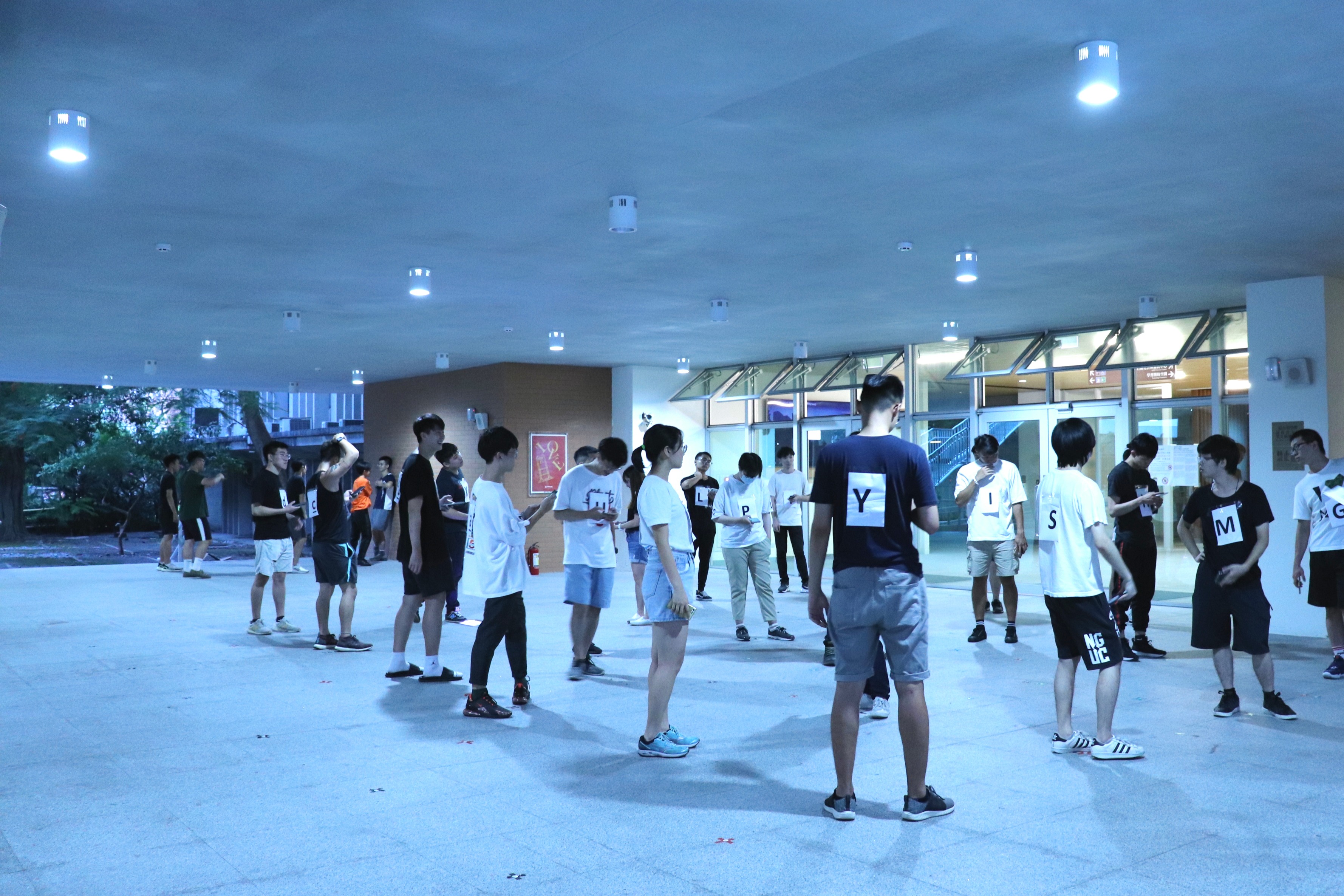}
         \caption{The outdoor scenario (a covered patio)}
         \label{fig:exp_pic_3}
     \end{subfigure}
        \caption{Settings and participants in the controlled experiment}
        \label{fig:exp_pics}
\end{figure*}

Our research design comprises four elements, as shown in Table~\ref{tab:methodology}. These are: 1) modification to the Covid-Watch-TCN mobile application; 2) a controlled experiment with 30 participants; 3) semi-controlled experiment with 50 participants in the field; and 4) a followup user survey administered to the participants from both experiments.
Research has been approved by REC Office at NTU (\#202006HS001).

% Our research design comprises four elements, as shown in Table \ref{tab:methodology}: A) the app customization, by adopting and modifying the Covid-Watch-TCN mobile application; B) controlled lab experiment with 30 participants; C) semi-controlled experiment with 50 participants in an actual event; and D) a following user survey for participants in B) and C).

% In the controlled experiment, we assign each participant a position and movement pattern, if any (see Figure~\ref{fig:instruction}\hc{is this the correct figure to show?} \js{Yes but the figure should be refined. @wj will find someone to do it.}), such that the distance between each pair of participants can be calculated. In the semi-controlled experiment, participants can move freely within a large auditorium, and the proximity events and contact events are partially reconstructed from video footage as well as on-site manual observation.

\subsection*{App Customization and Configuration}

To evaluate the effectiveness of Bluetooth-based decentralized contact tracing, our experiments collected information that would help us reconstruct the ground truth required for conducting comparative analysis: e.g., the sender of each token, and the distance between each pair of participants. Some of this information was collected by the mobile app and reported to a backend server; some was pre-assigned based on our protocol; and some was derived from direct observation. This subsection describes how we customized and configured an open-source app, and the following two subsections present our protocol and observational approaches, respectively.

% To evaluate the effectiveness of Bluetooth-based decentralized contact tracing, we collect additional information in our experiments to reconstruct the ground truth (e.g., the sender of each token and the distance between each pairs of participants), which is required for conducting comparative analysis. Some of the information is collected by a mobile app and reported to a backend server, some is pre-assigned based on our protocol design, and some by manual observation. We describe how we customize and configure an open-sourced app in this subsection, and present the protocol design and manual observation in the following subsections. 

We modified the source code of Covid-Watch-TCN Exposure Notification App~\cite{covidwatch}, whose Android and iOS versions are both open-source and can be found on GitHub. We modified the version based on the TCN Coalition's implementation. Since our use of it in our experiments, Covid-Watch has also now switched to using the GAEN APIs, which nevertheless was not enabled in Taiwan while we conducted this study.
Covid-Watch-TCN derives tokens (also called temporary contact numbers, TCNs) from a seed using cryptographic keys and hash functions. Keys are renewed periodically to balance storage overhead and privacy. While the detailed token generation algorithms, implementations, and parameter selection may vary, all of these apps rely on token reception and RSSI strength to estimate distance and exposure duration, and this is the main focus of our study.

To minimize interference with the app’s main functionality, we did not modify its token generation algorithm, where tokens are sent every 100 ms and changed every 15 minutes, but only locally logged data, and sent the logs back to our server at the end of each task. We manually inspected the app’s logic to identify the Android or iOS system API calls that created or sent tokens, and inserted our logging code before such calls.

On starting up, our modified app prompted each participant to enter the unique ID that was assigned to him/her at the beginning of the experiment, and to log device information including the running operating system and phone model.

While running, the app logged all the sent and received tokens, along with RSSI values, timestamps, and phone battery status. Specifically, when a device transmitted a token, the app logged the sent token with the unique ID of the device, the current battery status, and a timestamp. When a token was received from another participant’s device, on the other hand, the app logged that received token with measured RSSI and calculated distance; the unique ID of the device; and a timestamp. At the end of each task, the participants were asked to click on a “Submit” button to upload the logged information to our backend server. How the logged information was processed for further analysis will be explained in Section~\ref{sec:result}.

\subsection*{Phase 1: Controlled Experiment}
\begin{figure} [t]
\centering
\includegraphics[width=0.45\textwidth]{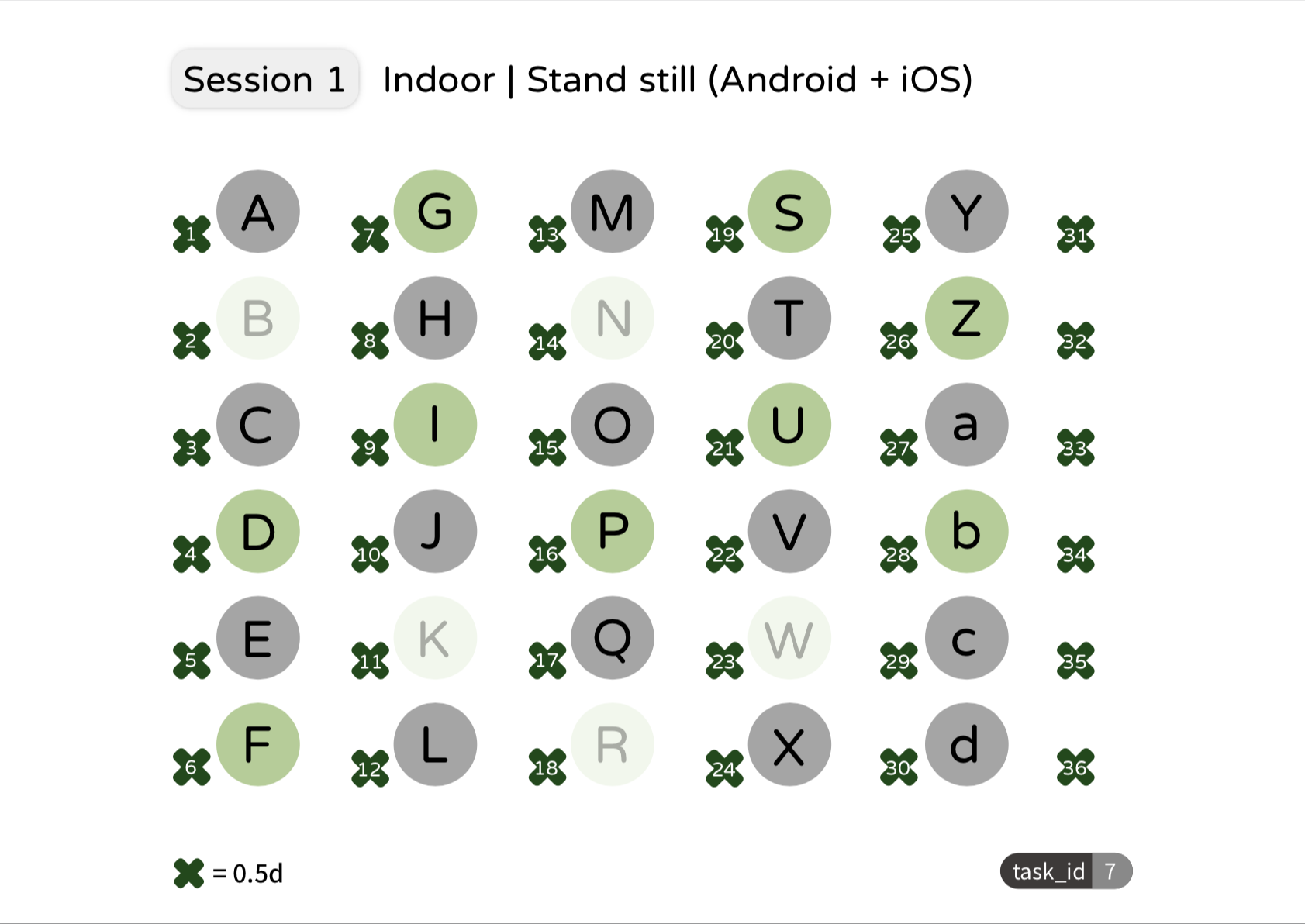}
\caption{Example of a slide showing participants' positions for a task. The letters represent users; the green circles, Android users; the black circles, iOS users; and the numerals, positions where each user stands.}
\label{fig:instruction}
\end{figure}

In Phase 1, each participant was assigned a position and movement pattern (Figure~\ref{fig:instruction}), such that the actual and Bluetooth-RSSI-estimated distances between each pair of participants can be calculated and compared at multiple time-points.

\paragraph{Research site and participants}

The controlled experiment, conducted in July 2020, comprised two scenarios, indoor and outdoor. The indoor scenario utilized an empty classroom $131m^2$ in size, and the outdoor scenario took place on a covered patio measuring $503m^2$. Both areas are $3.03m$ in height.

% indoor: 13.63 meters * 9.62 meters
% outdoor: 26.40 meters * 19.05 meters

We recruited 30 college students from our institutions, all of whom brought their personal mobile devices. Among these 30 devices, 14 were Android and the remainder were iOS. In both scenarios, we physically labeled each participant with a unique ID and marked the floor with tape. Before starting the experiment, we provided instructions to all participants explaining the research purpose and the overall experiment flow, and collected informed consent from all of them. Figure~\ref{fig:exp_pics} provides an overview of the experiment settings and process.

\begin{table} [t]
\caption{Session details of the pre-test. Note. $d = 1.5$ meters.}
\centering
\begin{tabular}{cccc}
    \toprule
    \textbf{Scenario} & \textbf{Task \#} & \textbf{OS(es)} & \textbf{Distance(s)} \\
    \hline
    \multirow{12}{*}{Indoor} & \multicolumn{3}{>{\columncolor[gray]{.8}}l}{Session 1: Stand still} \\
    %\cmidrule{2-4}
    & 1-3 & Android & \multirow{3}{*}{0.5d, 1.0d, 1.5d} \\
    & 4-6 & iOS & \\
    & 7-9 & Mixed & \\
    %\cline{2-4}
    & \multicolumn{3}{>{\columncolor[gray]{.8}}l}{Session 2: Stand still in a jammed environment} \\
    %\cline{2-4}
    & 10-12 & Mixed & 0.5d, 1.0d, 1.5d \\
    %\cmidrule{2-4}
    & \multicolumn{3}{>{\columncolor[gray]{.8}}l}{Session 3: Equidistantly walk in a given area} \\
    %\cmidrule{2-4}
    & 13-15 & Mixed & 0.5d, 1.0d, 1.5d \\
    %\cmidrule{2-4}
    & \multicolumn{3}{>{\columncolor[gray]{.8}}l}{Session 4: Gradually move closer} \\
    %\cmidrule{2-4}
    & 16 & Mixed & 0.5d \\
    %\cmidrule{2-4}
    & \multicolumn{3}{>{\columncolor[gray]{.8}}l}{Session 5: Stand still, and separated by a wall} \\
    %\cmidrule{2-4}
    & 17 & Mixed & 1.0d \\
    \hline
    \multirow{8}{*}{Outdoor} & \multicolumn{3}{>{\columncolor[gray]{.8}}l}{Session 6: Stand still} \\
    %\cmidrule{2-4}
    & 18-20 & iOS & \multirow{3}{*}{0.5d, 1.0d, 1.5d} \\
    & 21-23 & Android & \\
    & 24-26 & mixed & \\
    %\cmidrule{2-4}
    & \multicolumn{3}{>{\columncolor[gray]{.8}}l}{Session 7: Equidistantly walk in a given area} \\
    %\cmidrule{2-4}
    & 27 & Mixed & 1d \\
    %\cmidrule{2-4}
    & \multicolumn{3}{>{\columncolor[gray]{.8}}l}{Session 8: Gradually move closer} \\
    %\cmidrule{2-4}
    & 28 & Mixed & 0.5d \\
    \bottomrule
\end{tabular}
\label{tab:session_details}
\end{table}

\paragraph{Protocol}
The indoor scenario was broken down into five sessions, and the outdoor one into three, as shown in Table~\ref{tab:session_details}.
Both scenarios included three sessions---i.e., sessions 1, 3, and 4 in the indoor scenario, and sessions 6, 7, and 8 in the outdoor scenario---that required the participants to hold their devices in hands and i) stand still, ii) equidistantly walk in a given area, or iii) gradually move closer to each other. 

The other two sessions in the indoor scenario required the participants to stand still in a jammed environment characterized by continuous sending of useless data by Bluetooth beacons (Session 2), and near a wall that divided the participants into two groups (Session 5). To set up the jammed environment, we placed six RaspberryPi 3 model b devices in the same room. Each of these devices emitted unique, useless data via Bluetooth every 20 ms, i.e., five times faster than a normal device. This emulates two cases: 1) the existence of a malicious user jamming the wireless channels by sending tokens at a higher frequency, and 2) a very crowded place containing an additional 30 ($=6*5$) phones.

% The indoor and outdoor scenarios include five and three sessions, respectively. As shown in Table \ref{fig:session_details}, both scenarios contain three sessions (i.e., sessions 1, 3, 4 in the indoor scenario and sessions 6, 7, 8 in the outdoor scenario) that require the participants to hold their devices in hands and i) stand still, ii) equidistantly walk in a given area, or iii) gradually move closer to each other. There are two more sessions (i.e., sessions 2 and 5) in the indoor scenario that have the participants stand still i) in a noisy environment by continuously sending useless Bluetooth beacons in the background or ii) with a wall that divides the participants into two groups.

The eight sessions comprised 28 one-minute tasks. 
The multiple tasks within a session were set up to test how Bluetooth signal propagation varied across devices 1) running different operating systems (i.e., Android, iOS, and mixed) and 2) held at different distances from one another (i.e., 0.5d, 1d, 1.5d, where $d=1.5$ meters).

% The multiple tasks within a session were set up to test how the Bluetooth signal propagation varies by device 1) running different operating systems or 2) held at different distances.

\begin{itemize}
  \item \emph{Operating systems}. In sessions 1 and 6, where the participants stood still in both the indoor and outdoor scenarios, they were first grouped by their devices’ operating systems. After completing the tasks (i.e., tasks 1-6 and 18-23), the participants were taken out of these operating-system-based groups, and the experiment continued, any further groupings being randomized. 
  \item \emph{Distances}. To simulate a real-world setting in which people may or may not maintain social distancing, we asked the participants in sessions 1, 2, 3, and 6 to keep a distance of 0.5d, 1.0d, or 1.5d (d=1.5 m) from each other. This yielded data that subsequently allowed us to compare the estimated distances generated by the app against the ground-truth distances, and thus to evaluate the accuracy of the app’s proximity-detection techniques. Due to time limitations, sessions 4, 5, 7, and 8 were conducted with the participants attempting to maintain a single fixed distance of 0.5d, 1.0d, 1.0d, and 0.5d, respectively.
\end{itemize}

At the beginning of each task, the participants were prompted to turn on the app, as well as their devices’ Bluetooth and GPS. We required all users to turn on GPS during the experiment for consistency, because for Android version 6.0 and above, location services (e.g., GPS) need to be enabled when performing Bluetooth scanning~\cite{android6changes}. During each task, a series of slides would show the participants their assigned positions and/or movement paths (Fig.~\ref{fig:instruction}). Once a task ended, the participants were asked to manually upload the logged data to our backend server using the app.

\subsection*{Phase 2: Semi-controlled Experiment}
Unlike Phase 1, participants were allowed to move freely within a large auditorium during Phase 2. Proximity events and contact events were reconstructed based on video footage and direct on-site observation.

\paragraph{Research sites and participants}
The semi-controlled experiment was conducted at a summer school event in a campus auditorium measuring roughly $396m^2$ with a seating capacity of 250.

% Size of the NTUST IB-101 auditorium:
% (Roughly) 27.1 meters * 14.6 meters

Out of the 216 attendees, we recruited 50 participants: 26 were using Android devices and the other 24, iOS ones. The participants could be distinguished from the other attendees by differently-colored lanyards.

\paragraph{Protocol}
On the first day, we asked participants to turn on our research app, Bluetooth, and GPS for at least 90 minutes; on the second day, this was increased to 150 minutes. 
At the points when the participants were asked to turn on the app, they might be sitting still and listening to a speech, divided into groups and taking part in discussions, or having a tea break outside the auditorium.

One conference staff member was secretly assigned to be “the source of the virus” on the first day. This individual turned on the app, randomly passed by the participants, and recorded these actions with a GoPro camera so that we could reconstruct his close contacts after the experiment. Additionally, four researchers observed and manually documented the ground truth of proximity events in the auditorium, including the IDs of the participants involved and when they occurred.

\subsection*{User Survey}
After the experiments, we administered a questionnaire. Its three sections covered 1) technical problems the participants had encountered during the experiments, 2) their attitudes toward the use of the contact-tracing app, and 3) their attitudes toward personal privacy.

\paragraph{Section 1: Technical Problems}
In this section, the respondents could choose to agree with any or all of the following seven statements: \emph{Phone overheating, Seriously increased energy consumption, App crash, Unstable receiving token, Phone performance negatively affected, Couldn't log in}, and \emph{Other}. We also asked a yes/no question regarding whether the respondents had encountered upload failure during the experiments.

\paragraph{Section 2: Willingness to Use the Contact-tracing App} 
Section 2 aimed to capture how different technical factors and situations influenced the participants’ willingness to use the contact-tracing app. Its questions were divided into two groups.

In the first, each of the seven answer options from Section 1 regarding technical problems was repeated, along with the question, \emph{Will this technical problem affect your willingness to use the app in the next six months?} The respondent was then asked to select \emph{how much} each problem s/he had selected would affect such willingness, on a three-point Likert-scale ranging from \emph{-2=gradually decrease} to \emph{0=not influenced}, though an answer of \emph{N/A} could also be given in place of a scaled response.

The second group of items in Section 2 asked the respondents to select how various non-technical conditions would affect their willingness to use the app over the following six months. These conditions were \emph{Regulation by law or my school; Social influence from my family or colleagues; Planning a trip domestically or abroad; Entering a crowd of more than 100 people; and Current use of such apps by epidemic investigators}. These were rated on a five-point Likert-scale ranging from \emph{-2=gradually decrease} to \emph{+2=gradually increase}, plus an \emph{N/A} option.

\paragraph{Section 3: Privacy Considerations} In the final section of the questionnaire, the respondents were first asked about when and why they usually turned on their devices’ Bluetooth and GPS functions. Then, they rated the statement \emph{My data are secure and my privacy is protected while using the app} on a five-point Likert scale ranging from \emph{-2=strongly disagree} to \emph{+2=strongly agree}, again with an \emph{N/A} option. If a person’s response revealed a negative attitude toward the app’s privacy and security, i.e., was lower than 0, s/he would further be asked to select from among the following list of five data-security and privacy concerns: \emph{Data being tampered with; The app developer or associates may take advantage of security weaknesses; The app developer or associates may use my data for other purposes;} and \emph{My identity past contacts, or past locations may be recognized}.

We sent out the questionnaire to all 80 participants, but in fact this represented only 78 individuals, as two had participated in both experiments. Of these 78, 24 completed the questionnaire: a response rate of 30.8\%. 

\subsection*{Ethics}
This study was reviewed and approved in July 2020 by National Taiwan University’s Research Ethics Office (equivalent to an Institutional Review Board in North America), and meets all criteria for minimal-risk research (\#202006HS001).

\subsection*{Open Research Data}
We will open the participant instructions shortly after publication, so that other research teams can reuse our protocols or reproduce our research. 

\section*{Result}
\label{sec:result}

\subsection*{Data Processing and Analysis}

After removing incomplete data caused by tech glitches (5 Androids in the controlled experiment; 2 Androids and 7 iOS devices in the semi-controlled one were corrupted), the final dataset includes 66 devices.

Among the valid devices (33 Android and 33 iOS), Apple (n=33), Samsung (n=9), Google (n=5), and Xiaomi (n=5) were the most common models. Approximately half of the devices (n=30, 46.9\%) were existing models released within two years, and the remaining models were released two to five years before, as of July 2020. Forty-four devices (72.1\%) have been updated to Android 10.0, iOS 13.0, or newer versions released after September 2019. Thirteen (21.3\%) were Android 9.0 or iOS 12.1 onwards.

Based on the participants’ device logs, we reconstructed a directed multigraph, on which a vertex represents a participant and an edge from A to B represents a token sent by A and received by B. Each edge is labeled with a unique tuple (token, RSSI, timestamp) representing the corresponding token’s RSSI value and timestamp. Because tokens are sent every 100ms and changed every 15 minutes, the same token may be seen multiple times and have different RSSI values and timestamps. 

Tokens missing either sender or receiver information were removed. Missing sender information could have been caused by technical glitches (e.g., device malfunctioning, phone overheating and network congestion), while missing receiver information could have been caused by any of the same factors, or simply by no device having received them. Tokens with non-negative RSSI values or unrecognized sender/receiver IDs were also removed. In all, around 524,000 tokens, representing 86\% of the total received, were removed.

Because the participants all used their own devices, it was not possible for us to determine the root causes of all technical glitches; nor can we be certain that they were not specific to our experiments. However, app-store reviews and news reports reveal that many similar apps have struggled to resolve similar glitches in real-world settings. Thus, it seems relatively unlikely that our settings and/or app modifications caused them.

\subsection*{Phase 1: Results of Distance Estimation in a Controlled Environment}
The remaining 25 valid smartphones included nine Android and 16 iOS devices, and over the whole course of the experiment transmitted 700 unique tokens and received around 85,000 unique tuples of (token, RSSI, timestamp), all of which were included in the analysis described below.

Estimated distances between pairs of devices were calculated directly by the Covid-Watch-TCN app based on Bluetooth RSSI data.

The relation between measured RSSI and estimated distance, $d$, can be expressed as

\[RSSI = - 10n\log_{10}{d} + A\]

\noindent where $n$ is the environment factor, and $A$ is the reference signal strength at 1m. 

In both Android and iOS Covid-Watch-TCN apps, $n$ is set to 2.
The $A$ value is determined based on the sender's transmission power level, encoded in Bluetooth tokens. When receiving a token, the app extracts the transmission power level, and determines the value of $A$ according to what range that level falls within. Then, the app estimates based on the measured RSSI and $A$, using the equation shown above.

\subsubsection*{The Influence of Operating Systems}

\begin{figure} [t]
    \centering
    \includegraphics[width=0.4\textwidth]{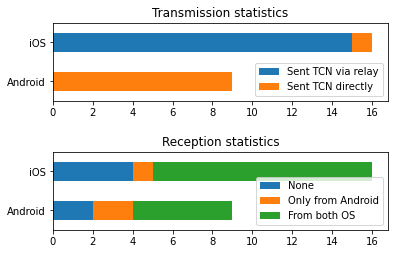}
    \caption{Transmission and reception statistics of Android and iOS devices}
    \label{fig:tx-rx-os}
\end{figure}

Figure~\ref{fig:tx-rx-os} represents the transmission and reception statistics for all the devices used during the controlled experiment, classified by operating system. It shows that there was a significant difference between Android and iOS devices' token-transmission capabilities.

Most of the Android devices were able to transmit tokens to both Android and iOS devices directly, while all but one of the iOS devices relied on nearby Android devices to broadcast tokens to others. However, two out of nine Android devices and four out of 16 iOS devices failed to receive any packets at all.

According to the Covid-Watch-TCN app’s specifications, iOS versions 13.4 and older do not support discoverability between third-party iOS apps in the suspended or background-running state if the devices’ screens are locked~\cite{covidwatch}. Therefore, they rely on Android devices as a relay to broadcast Bluetooth packets when running in the background. On the other hand, iOS devices running the app in the foreground exchange Bluetooth packets with one another directly.
Due to the iOS Bluetooth platform's reliance on relays from other devices, its RSSI and estimated-distance information cannot represent actual values. Therefore, we chose to focus only on directly transmitted tokens, i.e., Android-to-Android or Android-to-iOS, in our further analysis.

\subsubsection*{The Influence of Distance}

\begin{figure} [t]
    \centering
    \includegraphics[width=0.4\textwidth,]{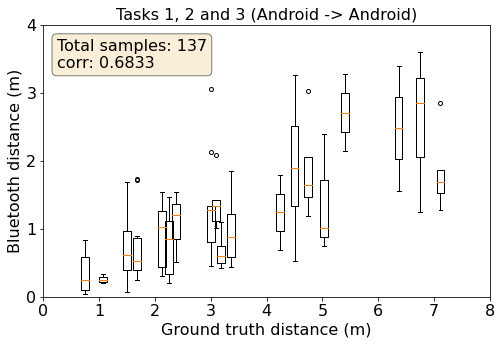}
    \caption{Estimated distance vs. ground truth, Session 1, Android phones only}
    \label{fig:indoor-android-boxplot}
\end{figure}

\begin{figure} [t]
    \centering
    \includegraphics[width=0.48\textwidth]{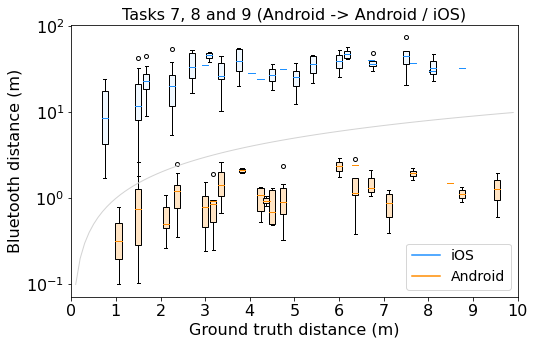}
    \caption{Estimated distance vs. ground truth, Session 1, phones with both systems running the app, both systems receiving}
    \label{fig:indoor-all-boxplot}
\end{figure}

Figure~\ref{fig:indoor-android-boxplot} illustrates the relationships of the estimated and true distances between each sender-receiver pair of Android devices in Session 1. In that session, the participants stood still in an indoor environment, and within each task were 0.5d, 1d, or 1.5d apart. The standard deviation of the estimated distance increased as the true distance increased, suggesting that Bluetooth signals attenuate during transmission and become more easily influenced by radio noise. The correlation coefficient between true distance and estimated distance is 0.68.

Figure~\ref{fig:indoor-all-boxplot} also indicates that the app tended to underestimate the distance between devices when the receiver was an Android one, potentially leading to high numbers of false-positive results. When the receiver was an iOS device, in contrast, the app tended to overestimate the distance, potentially leading to high numbers of false negatives.

\subsubsection*{The Influence of Background Radio Noise or Jamming}

\begin{figure*}[t]
     \centering
     \begin{subfigure}[b]{0.32\textwidth}
         \centering
         \includegraphics[width=\textwidth]{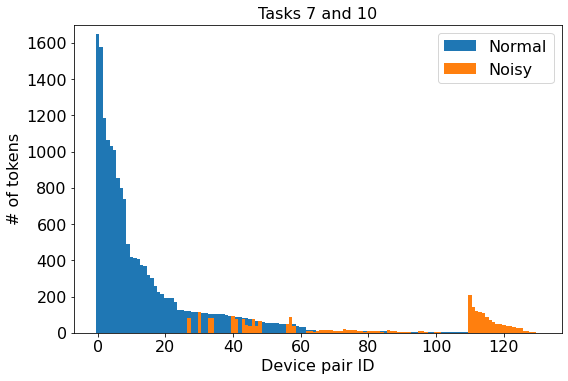}
         \caption{0.75m apart}
         \label{fig:sample-dif-0.5}
     \end{subfigure}
     \hfill
     \begin{subfigure}[b]{0.32\textwidth}
         \centering
         \includegraphics[width=\textwidth]{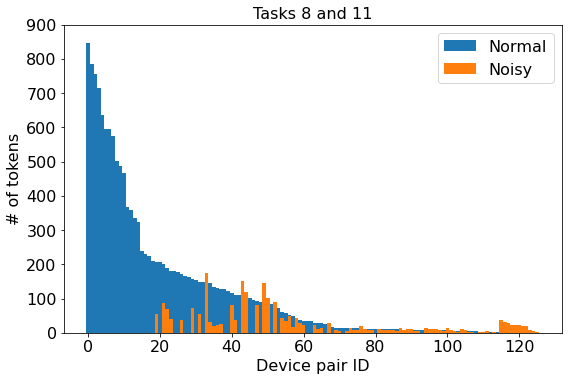}
         \caption{1.5m apart}
         \label{fig:sample-dif-1}
     \end{subfigure}
     \hfill
     \begin{subfigure}[b]{0.32\textwidth}
         \centering
         \includegraphics[width=\textwidth]{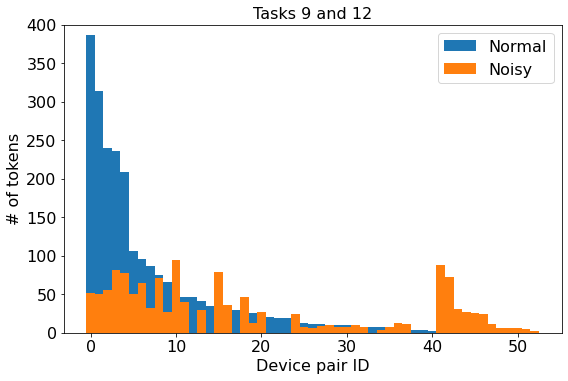}
         \caption{2.25m apart}
         \label{fig:sample-dif-1.5}
     \end{subfigure}
        \caption{Number of tokens received by each pair of devices in jammed vs. normal environments, with participants standing still}
        \label{fig:sample-diff}
\end{figure*}

Next, we tested if and how background radio noise or jamming affects token reception and RSSI.

Our results indicate that Bluetooth is susceptible to packet dropping due to jamming. 
The devices received fewer tokens when there were higher levels of background radio noise due to jamming. Figure~\ref{fig:sample-diff} shows that the number of received tokens was reduced in the presence of jamming. Additionally, the app had a lower accuracy when the participants were in a jammed environment. By applying a Wilcoxon Rank Test with $\alpha=0.05$, we confirm that the estimation errors across these two types of settings were drawn from two non-distinguishable distributions, implying that jamming did affect the app’s ability to estimate distance.

Even the app itself became a source of noise when a large number of app users were gathered in the same place. Tasks 7-9 can be seen as more ``noisy'' conditions than Tasks 1-3, i.e., with 16 iOS devices placed between each pair of Android devices. As shown in Figure~\ref{fig:indoor-all-boxplot}, as compared to the previous three tasks’ results, estimates of distance in ``noisy'' environments became more inaccurate in general, and even at distances of less than 2m. For these tasks, the correlation coefficient between true distance and estimated distance is just 0.26. 

Additionally, in Tasks 1-3, the app recorded transmission events for 137 out of 170 device pairs, a rate of 81\%. However, when more participants joined the experiment in Tasks 7-9, the number of recorded pairs dropped to 111, or 65\%; i.e., one-third of the senders were no longer able to successfully transmit tokens to receivers due to the “noise” caused by iOS devices in the immediate vicinity.

\subsubsection*{Power Consumption}
On average, the phone battery dropped by 11.3\% per hour in the uncontrolled experiment. 
We also observed a greater battery drop in larger crowds: the per-hour drops for small and large groups are 10.4\% and 29.6\%, respectively.

\subsection*{Phase 2: Results of Proximity and Contact Detection in a Real-world Event}
Data collected in the semi-controlled experiment were also analyzed to evaluate the effectiveness of the Covid-Watch-TCN app in a spacious indoor environment.
Collectively, over the two days of the second experiment, they transmitted a total of 39,000 tokens and received 1.8 million.

A \emph{proximity event} was deemed to have occurred if 1) two devices were detected by the app as having exchanged tokens at below a particular estimated-distance threshold, and 2) the time at which this exchange was recorded as occurring by the app was within 15 minutes before or after the time at which the same event was recorded by the researchers observing the conference and/or the GoPro videos.

A \emph{contact event} between two devices was defined as a continuous proximity event lasting for a particular period, for example, 15 minutes. We further defined a \emph{strict contact event} as one meeting the additional condition that every minute during the exposure period includes at least one proximity event.

\subsubsection*{Proximity Detection and Contact Tracing}

\begin{figure} [t]
    \centering
    \includegraphics[width=0.4\textwidth]{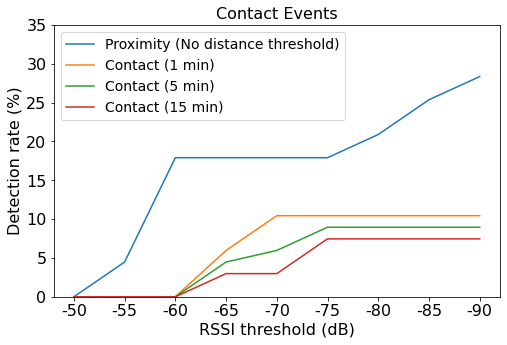}
    \caption{Detection rates for contact events of three durations by RSSI threshold, with proximity detection shown for comparison}
    \label{fig:proxy-cont-rate}
\end{figure}

\begin{figure} [t]
    \centering
    \includegraphics[width=0.4\textwidth]{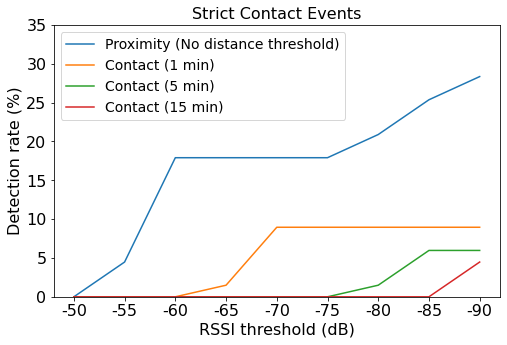}
    \caption{Detection rates for strict-contact events of three durations by RSSI threshold, with proximity detection shown for comparison}
    \label{fig:proxy-cont-rate-strict}
\end{figure}

There were 67 proximity events documented by the four researchers during the experiment and extracted from the GoPro videos. If we set the distance threshold as 2m (commonly recommended as a social-distancing measure), only 16 of these proximity events were detected, implying a proximity detection rate of 24\%. However, even when no distance threshold was set (i.e., no lower bound was placed on the RSSI value), the number of detected events only increased to 19, i.e., 28\% of the total known to have occurred.

Under an exposure-duration rule of 15 minutes, meanwhile, the app could only detect 7.5\% of the relevant contact events. Decreasing the exposure duration to 5 minutes and 1 minute resulted in only slight increases in the contact-detection rate: to 9.0\% and 10.4\%, respectively. Additionally, when the strict contact rule was applied, the app failed to detect any contact events at all. The proximity and contact detection rates at various RSSI and contact-duration thresholds are shown in Figure~\ref{fig:proxy-cont-rate} and Figure~\ref{fig:proxy-cont-rate-strict}, in which a measured signal with RSSI of -80 dB equates roughly to a 2m separation.

We also evaluated the proximity and contact detection rates of ``the source of the virus''. A total of 11 proximity events (lasted for at least 5 minutes) with this individual were recorded, two via direct observation and nine via review of the GoPro videos. However, only four of these 11 proximity events were recorded by the app, despite none being fleeting. That is, exposure to the “virus” lasted for at least 5 minutes in each case, according to our observations. Moreover, among the 11 documented 5min-contact events involving the “virus”, none were detected. 

\subsubsection*{Exposure Duration}
\begin{figure} [t]
    \centering
    \includegraphics[width=0.4\textwidth]{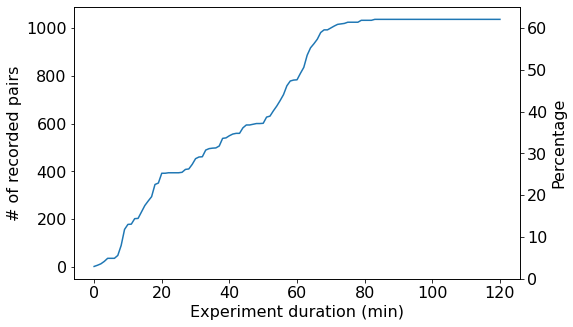}
    \caption{Change over time in the number of unique recorded sender-receiver pairs}
    \label{fig:pair-count}
\end{figure}

\begin{figure} [t]
    \centering
    \includegraphics[width=0.4\textwidth]{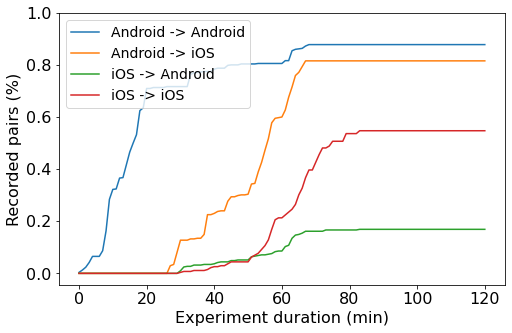}
    \caption{The percentage of recorded unique sender-receiver pairs classified by the running operating system}
    \label{fig:pair-count-os}
\end{figure}

During the span of the semi-controlled experiment, each device sent at least one token to every other device, and received at least one, with an average of around 1,000 tokens per device being sent, and about 38,000 per device being received.

Figure~\ref{fig:pair-count} illustrates trends in the number of unique recorded sender-receiver pairs over a 90-min period on the first day of the experiment.  This number steadily increased over time and converged to an upper bound of about 1,000, with 63\% of all possible pairs represented. This indicates that, if one of its users remains in an indoor environment long enough, the app will be able to discover most nearby devices.

We further classified the recorded device pairs according to their operating systems, as shown in Figure \ref{fig:pair-count-os}. Most (> 80\%) of the Android devices were discovered by both nearby Android and iOS devices. As for iOS devices, 55\% were discovered by iOS devices, with only 17\% discovered by Android devices. The results are consistent with our findings in the controlled experiment that limitations of the iOS Bluetooth platform could significantly influence the transmission capability of iOS devices. 

\subsection*{Users’ Perceptions of App Usage}

Among the 24 questionnaire respondents, about half of them indicated that they had encountered at least one technical problem, including \emph{the app crashing}(n=7), \emph{unstable receiving tokens} (n=6), \emph{phone overheating} (n=4), \emph{unexpectedly high energy consumption} (n=3), \emph{login issues} (n=3), and \emph{difficulty uploading their devices' token data} (n=15). Nine out of the 15 participants tried re-uploading and successfully uploaded the data eventually.

Among these, the technical problem with the strongest negative influence on the respondents’ intentions to use the app was high energy consumption, followed by phone overheating. We also found that, although marginally more respondents mentioned experiencing \emph{crash problems} (n=13) than \emph{inefficient phone performance} (n=12), the latter problem had a greater negative impact on their willingness to use the app.

The external conditions that the respondents selected most often as likely to affect their willingness to use the contact-tracing app were \emph{entering crowds of 100 people or more} (n=19); \emph{current use of the app by epidemic investigators} (n=19); \emph{regulations} (n=16); and \emph{domestic-trip planning} (n=16). The top two of these conditions were also the most positively influential on the respondents’ willingness to use the contact-tracing app. Some of the participants even rated the influence of regulations on their willingness to use the app as negative. 

Turning now to privacy issues, half our respondents stated that their habits regarding Bluetooth and GPS functions would not change in the wake of our experiments, whereas half said that they would. However, since our questionnaire did not ask about how/why Bluetooth and GPS usage impacted the respondents’ privacy concerns, we cannot make any conclusions about this split in attitudes.

Surprisingly, only a small minority of our respondents expressed a belief that, due to using the focal contact-tracing app, their \emph{data} (n=4) or \emph{privacy} (n=5) might be unsafe, with the others either deeming them to be safe, or expressing no opinion on this matter. Among the minority, the top two data-security concerns cited were that \emph{the app developer might take advantage of security weaknesses} (n=4), and that \emph{the app developer did not build in sufficient protections} (n=3). Their top privacy concerns included \emph{their past routes being recognized} (n=4), \emph{developers' associations using the data for other purposes} (n=4), and \emph{the developer itself using the data for other purposes} (n=3). 
\section*{Practical Implications}
\label{sec:discussion}

Our empirical findings have four important implications for \BDCT, discussed in turn below.

\subsection*{Reliable or Unreliable Proximity Detection?}
RSSI alone does not produce reliable estimates of physical distance, which aligns with the findings of previous studies in indoor positioning~\cite{BCCGD16} and contact tracing~\cite{leith2020coronavirus}. In our experiments, the RSSI estimates often spanned $-11.3$ to $11.7$ dB, resulting in errors 0.27 to 3.85 times of the ground truth. Unreliable distance estimates lead to inaccurate proximity or contact detection. 

While increasing the number of samples might reduce this variance, we observed system bias caused by contextual factors such as phone models and crowd size, in addition to those investigated in previous work, including wall geometry, phone orientation, and whether users were indoors or outdoors. These system biases would be difficult to eliminate in the absence of extensive, detailed prior knowledge of the context in which contact tracing would need to occur. 

Although the app was unreliable in estimating the distance between app users, information about whether they are in the same indoor location or not could still be useful to the broader contact-tracing process.

\subsection*{Influence of Crowds and Jamming}
Another observation was that variance increased with the density of the crowd.
The variance was lower when the participants were farther away from each other. This could be due to the status of human bodies as obstacles and wireless channels becoming congested when all devices in the room are transmitting signals simultaneously.
In addition, the six Raspberry Pi devices that emitted tokens at a high rate in one session of our experiment, which we added to investigate possible jamming effects, had a similar impact on more crowded conditions. All else being equal, the variance was higher in the more ``noisy'' environment that resulted from the inclusion of these extra devices.

Reducing the token broadcast frequency (e.g., from 100ms to 1s) in dense areas may alleviate packet loss due to interference, but its effect on the proximity and contact detection remains to be investigated.

\subsection*{Interoperability Issues}
To be effective, apps need to be interoperable and produce consistent results regardless of what OSs, phone models, app configurations, and implementations are involved.  Our experiments used the existing Android and iOS versions of the same app, and about half of our participants used Android, and the rest used iOS. To emulate realistic scenarios, we did not limit the phone models involved, apart from a requirement that all must support Bluetooth.

We found asymmetric results across phone models and versions. The differences we observed among phone models might be due to differences in Bluetooth chips, transceiver modules, and signal-processing methods, among other factors. This complicates interoperability by implying that each receiving phone may need to know the model and version of each sending phone if the app’s detection accuracy is to be improved.

We observed that iOS devices tended to overestimate distances, while Android ones tended to underestimate them. The overestimation by iOS devices may have been caused by the calibration of the default reference RSSI value (i.e., $A$) at 1m across both versions of the Covid-Watch-TCN app.
For the Android version, there were only three reference values; the iOS version had the same possible values as the Android version, except that its default value was greater by 10, i.e., Android is -67db and iOS -57db. These coarse ranges of transmission power levels could lead to inaccuracy in distance estimates.

GAEN system~\cite{ExposureNotification} recommends that Android devices be calibrated to a typical iPhone according to their model designations. However, even with improved calibration to compensate biases due to inter-device differences, inaccuracies caused by environmental factors may be difficult to eliminate in the absence of prior contextual knowledge.

\subsection*{Hidden Issue---Power Drain and Other Glitches}
In all our experimental tasks, participants’ phone batteries drained quickly regardless of brand. This excessive consumption means that our research app would not be usable in real-world scenarios, even if people were willing to try. Some also complained that their phone overheated while running the app.

This extreme power consumption may be attributable to how our app handles Bluetooth. The early version of Covid-Watch---along with many other apps implemented before the release of the GAEN API---did not have native Bluetooth access and had to use hacks to bypass low-level restrictions. For example, iOS versions below 13.4 can only send tokens when either the sender or the receiver is running in the foreground. These hacks likely consume unnecessary resources, including energy. Although we were unable to test GAEN API-based apps, we anticipate that they will have better power efficiency.

Some participants also experienced app crashes or hangs, and thus could not broadcast or submit tokens. Although this was likely caused by our rapid development cycle and lack of testing on a variety of phone models and OSs, it is worth noting that similar issues have been reported by users of other contact-tracing apps.

\subsection*{Users' Perception}
Our survey result may indicate that people's willingness to use contact-tracing apps is rooted in self-protection concerns and/or a public-spirited desire to aid epidemic investigation work, but that the enforced use of such apps might nevertheless provoke opposition. 

However, it should be borne in mind that most of our participants in the second experiment were students with information-engineering backgrounds and an interest in cybersecurity, who may have been less likely to worry about data-security issues than an equivalent-sized sample of the general public.

\subsection*{Limitations}
\label{ssec:limitations}

Additional studies are needed to address the following limitations:
1) Our evaluation was restricted to a specific implementation. Using the GAEN APIs might alleviate battery and interoperability issues.
2) Our logging code may introduce additional overhead.
3) Our experiments were of short duration and not representative of the full range of real situations.
4) Falsely identified non-contacts (false positives) were not analyzed in our semi-controlled setting.

\section*{Acknowledgments}
This work was financially supported by the Ministry of Science and Technology (MOST) in Taiwan, under MOST 107-2221-E-009-028-MY3, 109-2636-E-002-021-, and 109- 2636-H-002-002-.
The authors would like to thank Xue-Yuan Gu and Kai-Lin Zhang for co-creating the application, Jia-Chi Huo and Bo-Rong Chen for implementing the jamming devices, Yun-Chi Chang, Yu-Jen Chen, Li-Fei Kung, and Yu-Ju Yang for helping conduct the experiments, and Yu-Wen Huang for redesigning the instruction slides.

\nolinenumbers
\bibliography{paper}

\end{document}